\documentclass[aps,prl,twocolumn,floatfix,superscriptaddress]{revtex4}
\usepackage{graphicx,amsmath,amsfonts}
\usepackage{textcomp}
\usepackage{units}
\usepackage{epsfig}
\newcommand{\0}{|0\rangle}
\newcommand{\1}{|1\rangle}

\newcommand{\yb}{Yb$^+$\,}

\newcommand{\Dhalf}{D$_{3/2}$}
\newcommand{\jK}{$[3/2]_{1/2}$}

\begin{document}
\graphicspath{{Bilder/}}

\title{Error-resistant Single Qubit Gates with Trapped Ions}

\author{N. Timoney}
\affiliation{Fachbereich Physik, Universit\"at Siegen, 57068 Siegen,
Germany}
\author{V. Elman}
\affiliation{Fachbereich Physik, Universit\"at Siegen, 57068 Siegen,
Germany}
\author{W. Neuhauser}
\affiliation{Institut f\"ur Laser-Physik, Universit\"at Hamburg,
Luruper Chaussee 149, 22761 Hamburg, Germany}
\author{Chr. Wunderlich}
\affiliation{Fachbereich Physik, Universit\"at Siegen, 57068 Siegen,
Germany} \email{wunderlich@physik.uni-siegen.de}
\date{\today}

\begin{abstract}
Coherent operations constitutive for the implementation of single
and multi-qubit quantum gates with trapped ions are demonstrated
that are robust against variations in experimental parameters and
intrinsically indeterministic system parameters. In particular,
pulses developed using optimal control theory are demonstrated for
the first time with trapped ions. Their performance as a function of
error parameters is systematically investigated and compared to
composite pulses.
\end{abstract}


\maketitle

In order to experimentally implement a device capable of performing
fault-tolerant universal quantum computation (QC), quantum gate
operations involving one or multiple qubits have to be carried out
with demandingly high accuracy (see, for instance,
\cite{Aliferis06,Knill06}). According to recent theoretical
investigations, the experimentally required accuracy of quantum
gates for fault-tolerant universal quantum computation no longer
seems daunting or even prohibitive \cite{Knill06}. But still, the
desired error probability per gate (EPG) should be as small as
possible in order to keep the experimental overhead necessary for
quantum computation within a feasible limit. Thus a low error
probability is prerequisite for scalable fault-tolerant QC.

Any quantum algorithm can be decomposed into a sequence of unitary
operations applied to individual qubits (single-qubit gate) and
conditional quantum dynamics with at least two qubits
\cite{Barenco95}. Multi-qubit gates (involving two or more qubits)
are synthesized by applying a sequence of elementary unitary
operations on a collection of qubits. Each of these elementary
operations is often similar, or identical, to what is needed for
single-qubit gates, and therefore each operation has to be
implemented with an error probability well below the tolerable EPG
characterizing the full gate operation.

If electrodynamically trapped ions are used as qubits, then a
unitary operation amounts to letting ions interact with
electromagnetic radiation with prescribed frequency, phase,
amplitude, and duration of interaction in order to implement quantum
gates. Recently, impressive experimental progress was demonstrated
in entangling up to eight ions, and performing 2-qubit quantum gates
\cite{Leibfried05,SchmidtKaler03,Leibfried03}. Architectures
allowing for scalable QC with trapped ions have been proposed ({\it
e.g.}, \cite{Kielpinski02}), and building blocks necessary for
achieving this ambitious goal are currently being investigated using
various types of ions.

The error budget, for instance, of the geometrical phase gate
demonstrated in \cite{Leibfried03} is dominated by the frequency and
amplitude uncertainty of the laser light field. These errors are
also responsible for a part of the EPG of the controlled-NOT gate
reported in \cite{SchmidtKaler03}. If an "ion spin molecule", that
is, trapped ions coupled via a long range spin-spin interaction, is
to be used for quantum information processing, then the exact
transition frequency of a particular ionic qubit depends on the
internal state of other ions \cite{Wunderlich02}. Therefore, here
too, it is important to have quantum gates at hand that are
insensitive to the detuning of the radiation driving the qubit
transition.

Here, we demonstrate single qubit gates
with trapped ions that are robust against experimental
imperfections over a wide range of parameters. In particular it is
shown that errors caused by an inaccurate setting of either
frequency, amplitude, or duration of the driving field, or of a
combination of these errors are tolerable (in terms of a desired
accuracy of quantum gates) when a suitable sequence of radiation
pulses is applied instead of, for instance, a single rectangular
$\pi$-pulse. Thus an essential prerequisite for scalable quantum
computation with trapped ions is demonstrated.

We shall show results of using shaped pulses which were developed
using optimal control theory. Optimal control theory relies on a
generalization of the classical Euler-Lagrange formalism. Using
discrete rather than continuous trajectories in the configuration
space of the quantum mechanical system, and constraining the
trajectories to satisfy the Bloch equations, the cost function to
be optimized, is the transfer efficiency over a range of frequency
detunings between qubit and applied radiation and amplitudes of
the radiation \cite{Skinner03}. In addition, composite pulses are
realized here that are specifically designed to tackle
off-resonance errors, or designed to tackle pulse length errors or
power discrepancies of the driving field \cite{Cummins03}.

The two level quantum mechanical system used as qubit is realized
on the $\0\equiv$ S$_{1/2}$(F=0) $\leftrightarrow$ S$_{1/2}$(F=1,
m$_{F}$=0) $\equiv \1$ transition in a single $ ^{171}$Yb$^{+}$
ion with Bohr frequency $\omega_0$  confined in a miniature Paul
trap (diameter of 2 mm), driven by microwave radiation with
frequency $\omega$ close to $2\pi \times 12.6$ GHz and Rabi
frequency $\Omega \approx 2\pi\times 10$ kHz. The time evolution
of the qubit is virtually free of decoherence, that is,
transversal and longitudinal relaxation rates are negligible, and
is determined in a rotating frame after the rotating wave
approximation, by the semi-classical Hamiltonian
 \mbox{$
 H= \frac{\hbar}{2} \delta \sigma_z
    + \frac{\hbar}{2} \Omega  \left( \sigma_+ e^{-i\Phi} + \sigma_-
    e^{i\Phi}\right) \ .
 $}
Here, $\sigma_{\pm}$ are the atomic raising/lowering operators,
$\sigma_z$ is a Pauli matrix, and $\delta\equiv \omega_0 - \omega$
is the detuning of the applied radiation with respect to the
atomic transition. Imperfect preparation of the $\0$ state by
optical pumping, limited here the purity of the initially prepared
state, such that the initial density matrix (before coherent
interaction with microwave radiation) is given by $\rho_{i}=a_1 \1
\langle 1| + a_0 \0 \langle 0|$ with typical values $a_1 = 0.1$
and $a_0 = 0.9$.

The ion is produced from its neutral precursor by photoionization
using a diode laser operating near 399 nm. Laser light near 369 nm
driving resonantly the S$_{1/2}$ F=1 $\leftrightarrow$ P$_{1/2}$
F=0 transition in \yb is supplied by a frequency doubled Ti:Sa
laser, and serves for cooling and state selectively detecting the
ion. A diode laser delivers light near 935 nm and drives the
\Dhalf - \jK transition to avoid optical pumping into the
metastable D$_{3/2}$ state during the cooling and detection
periods (for details see \cite{Wunderlich03}). For quantifying
errors in detuning we will use the scaled detuning, $f=\delta /
\Omega$ whereas errors in pulse area shall be represented by $g =
\Delta\theta/\theta $, where $\Delta\theta = \theta' - \theta$
with the desired pulse area, $\theta = \int_0^T\Omega dt$ and the
actual pulse area, $\theta'$ when $T$ or $\Omega$ are not set
perfectly. The fidelity of the qubit state
 \mbox{$
 |\theta_m,\phi_m\rangle \equiv \cos\left(\theta_m/2\right)\mid 0\rangle+
e^{(i\phi_m)}\sin\left(\theta_m /2 \right)\mid 1\rangle
 $}
that is obtained after applying a microwave pulse is given by
 $
F = \mid \langle \theta, \phi\mid \theta_m,\phi_m \rangle \mid^2
 $
with $|\theta,\phi\rangle$ being the state that would be obtained,
if the pulse were perfect. Thus, for a $\theta=\pi$-pulse the
fidelity is given by
 $
F_{\pi} = \mid\sin\left(\theta_m /2 \right)\mid^2
 $.
Impure initial preparation ($a_1>0$) limits the maximum fidelity
that can be obtained with a $\pi$-pulse to $ F_{\pi} = a_0 - a_1$.
In order to determine the fidelity of the state obtained after a
gate that is supposed to leave the qubit with a well defined phase
$\phi$ ({\it e.g.} a $\pi/2$-pulse), the phase $\phi_m$ needs to
be determined in addition to $\theta_m$. A Ramsey-type experiment
allows for measuring both angles: First an ideal Ramsey sequence
is carried out (i.e. two successive ideal $\pi/2$-pulses with
varying phase $\Phi$ of the second pulse) yielding interference
fringes in the population of state $\1$, $a_1$ as a function of
$\Phi$. Then, the first $\pi/2$-pulse is replaced by a possibly
non-ideal pulse sequence leaving the qubit in state
$|\theta_m,\phi_m\rangle$, and again interference fringes, are
recorded. Now the population $a_1(\Phi)$ detected in state $\1$ is
given by
 \mbox{$
a_1(\Phi)=1/2(1+ \sin(\theta_m)\cos(\phi_m-\Phi))
 $},
and from a fit of the data one obtains $\theta_m$ and $\phi_m$.

The basic measurement sequence (labeled sequence A) for
determining the fidelity as a function of $f$ and $g$ of a shaped
pulse that ideally gives a rotation with $\theta=\pi$ is as
follows: i) A single ion is prepared in the $\0$ state by optical
pumping through illumination with 369 nm light for 20ms. ii) A
microwave shaped pulse with controlled error, that is, known
values of $f$ and $g$, is applied. iii) Again, the ion is
illuminated for 4 ms with 369 nm light for state selective
detection. iv) The ion is laser cooled by applying microwave and
laser radiation simultaneously. This sequence comprising steps i)
through iv) is then repeated (sequence B), except that in step
ii), for direct comparison, the shaped pulse is replaced by a
rectangular pulse that would give $\theta=\pi$, if $f=0=g$. Then,
i) through iv) is repeated again (sequence C) with an ideal ({\it
i.e.}, $f=0=g$) $\pi/2$ rectangular pulse to control state
selective detection in each measurement cycle. Finally, i) through
iv) is repeated a fourth time (sequence D) leaving out step ii) in
order to monitor the initial preparation in terms of the
coefficient $a_0$. Typically, the full procedure (sequences A
through D) outlined here is repeated 700 times for a given pair of
$f$ and $g$ values.

\begin{figure}
\centering
      {\includegraphics[width=.35\textwidth]{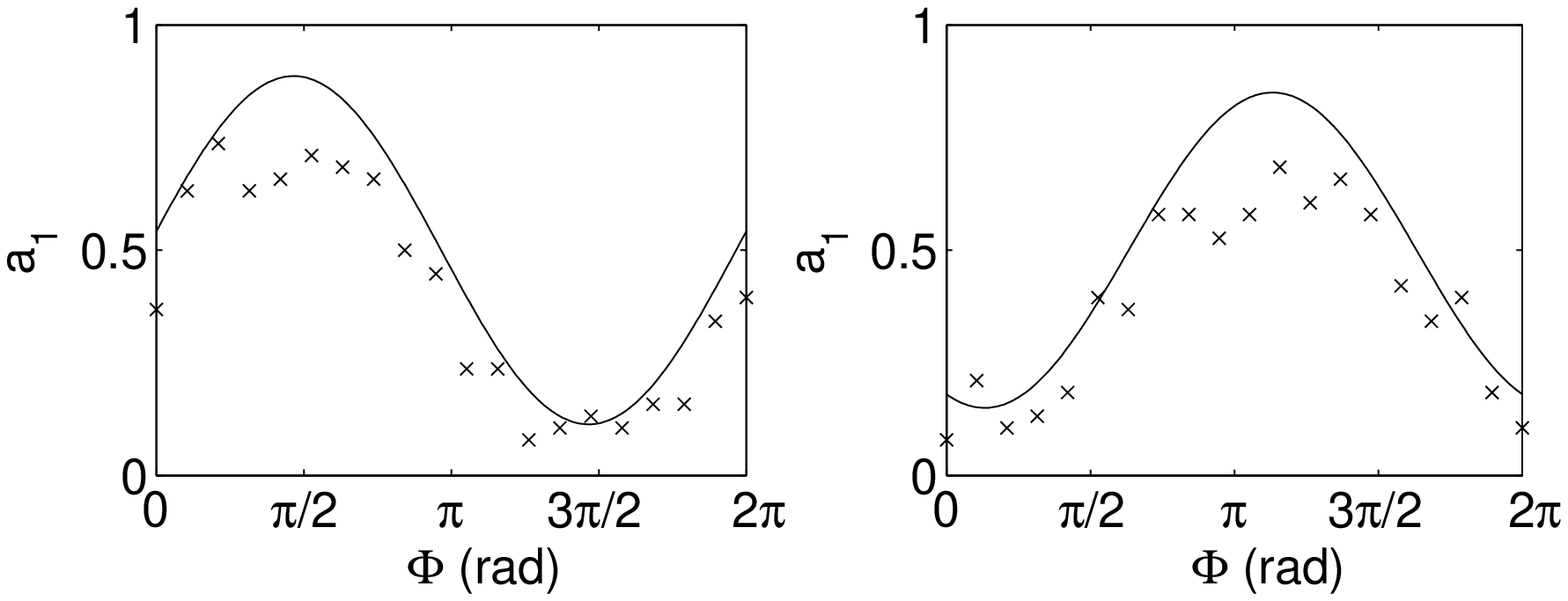}}
      {\includegraphics[width=.35\textwidth]{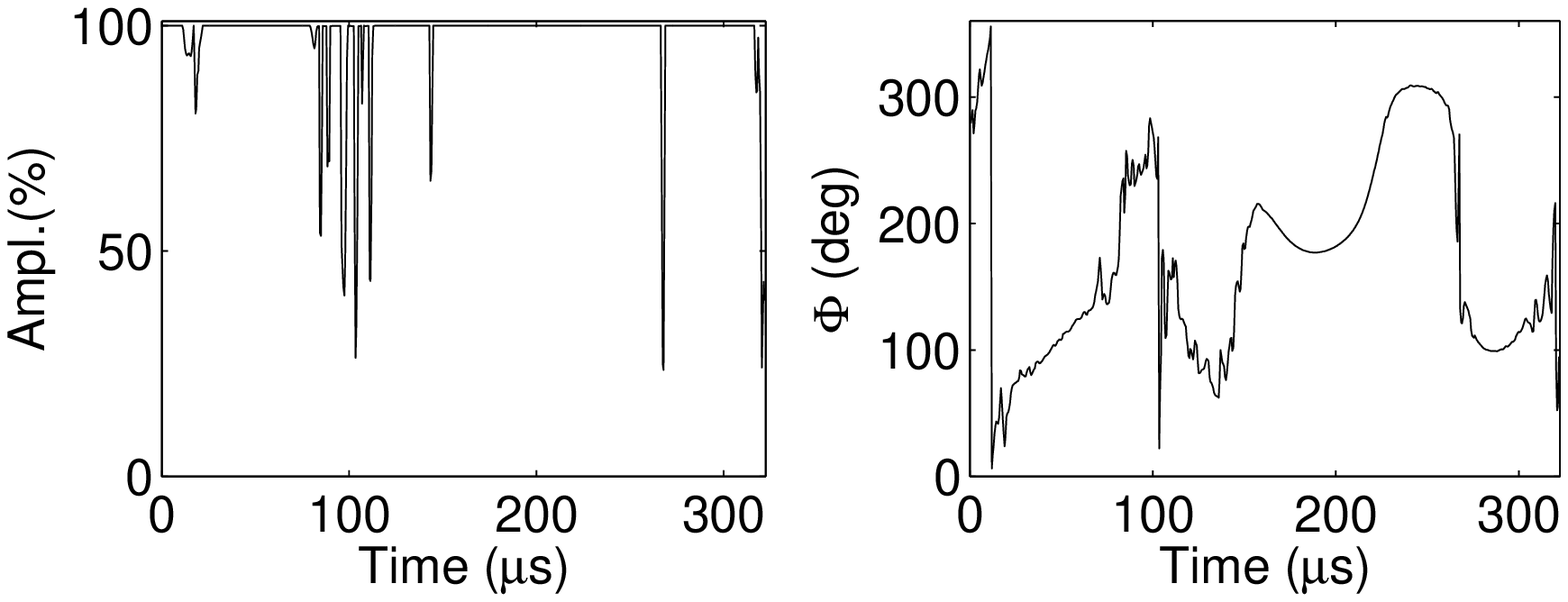}}
\caption{Upper graphs: Ramsey interference fringes from sequence
A' (left hand side) and sequence B' (rhs, $f=-1$, $g=0$). A fit
(indicated by a solid line) of the experimental data (crosses)
gives $\theta_m=3.92 \pm 0.19, \phi_m= 0.418\pm 0.34$ (see text).
Lower: Relative amplitude (lhs) and phase (rhs) of a pulse
consisting of 645 individual steps evolving over time. This pulse
was developed using optimal control theory \cite{Luy04}.}
    \label{fig:Ramsey_Fit}
\end{figure}

When measuring the performance of shaped pulses and pulse
sequences that ideally yield a rotation of $\theta=\pi/2$ and, for
instance, $\phi= -\pi/2$, then the basic sequence, A' is as
described above, except that in step ii) two ideal ($f=0=g$)
$\pi/2$ pulses are applied, the second one with variable phase
$\Phi$ (this yields Ramsey fringes as a reference). Then, sequence
B' is performed by replacing the first $\pi/2$ pulse in step ii)
by a shaped pulse with controlled error. Sequence C' is obtained
by replacing the first $\pi/2$ pulse in ii) with a rectangular
pulse subject to the same errors as the shaped pulse in B'.
Sequences A' through C' are repeated 20 times while increasing the
value of $\Phi$ by $2\pi/20$. Then sequences C and D are appended
and the complete procedure is repeated 50 times. As an example,
Fig. \ref{fig:Ramsey_Fit} upper panels illustrates how angles
$\theta_m$, $\phi_m$ were determined for a specific pair of $f$
and $g$ values.

A completed measurement returns two grids of fidelities with the
points on each grid defined by different values of $f$ and $g$.
One grid corresponds to a simple pulse (obtained from sequences B
or C'), that is, the amplitude as a function of time has a
rectangular shape with the desired pulse area, $\theta=\Omega t$.
This pulse results in a perfect rotation by an angle $\theta$ only
for $f=0$ and $g=0$. The second grid of fidelities corresponds to
a shaped or composite pulse (obtained from sequence A or B'). In
the experimental results shown below, crossing points between
gridlines represent measurement points. The shaded areas are
obtained by linear interpolation between points.

The pulses developed using Optimal Control Theory and demonstrated
here were designed for off-resonant errors described by $-1 < f <
1$, power variations up to $\pm 40\%$, Rabi frequency $\Omega =
2\pi\times 10$ kHz, and are made of $0.5\mu$s steps. In each step,
in general, amplitude and phase of the radiation is changed. As an
example the time evolution of phase and amplitude of a shaped
pulse consisting of 645 individual steps is shown in Fig.
\ref{fig:Ramsey_Fit} \cite{Luy04}. This pulse is supposed to yield
a rotation of $\theta =\pi/2, \phi=-\pi/2$ that is robust against
variations in $f$ and $g$.

\begin{figure}
     {\includegraphics[width=.4\textwidth]{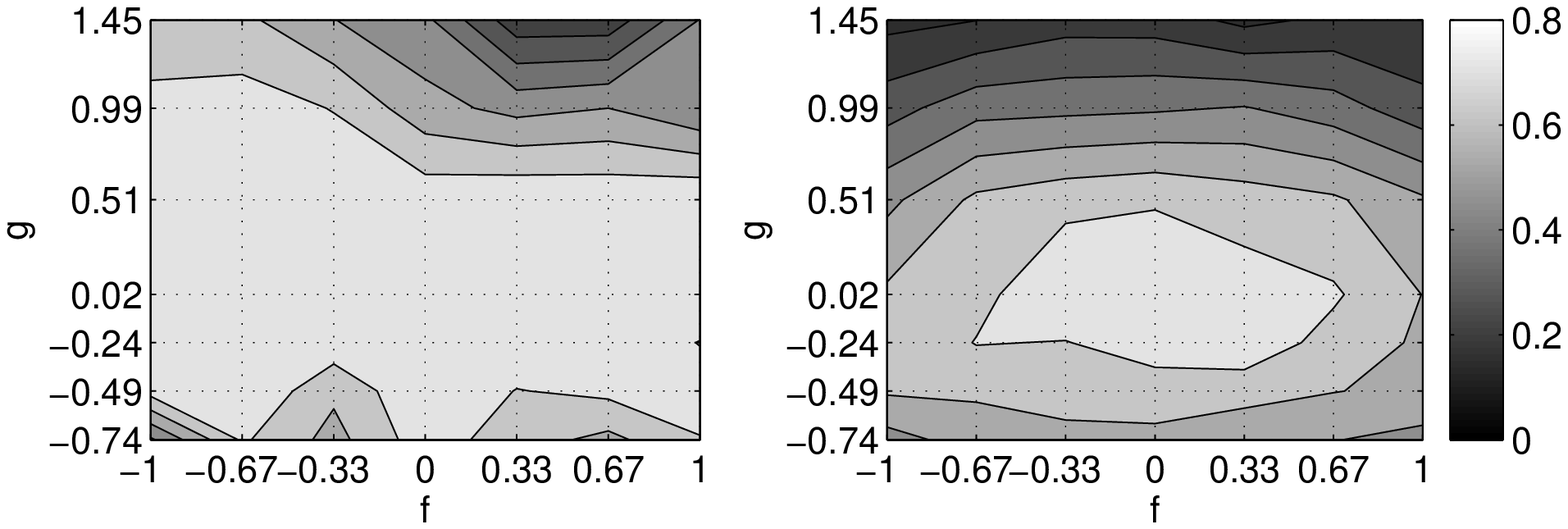}}
     {\includegraphics[width=.125\textwidth]{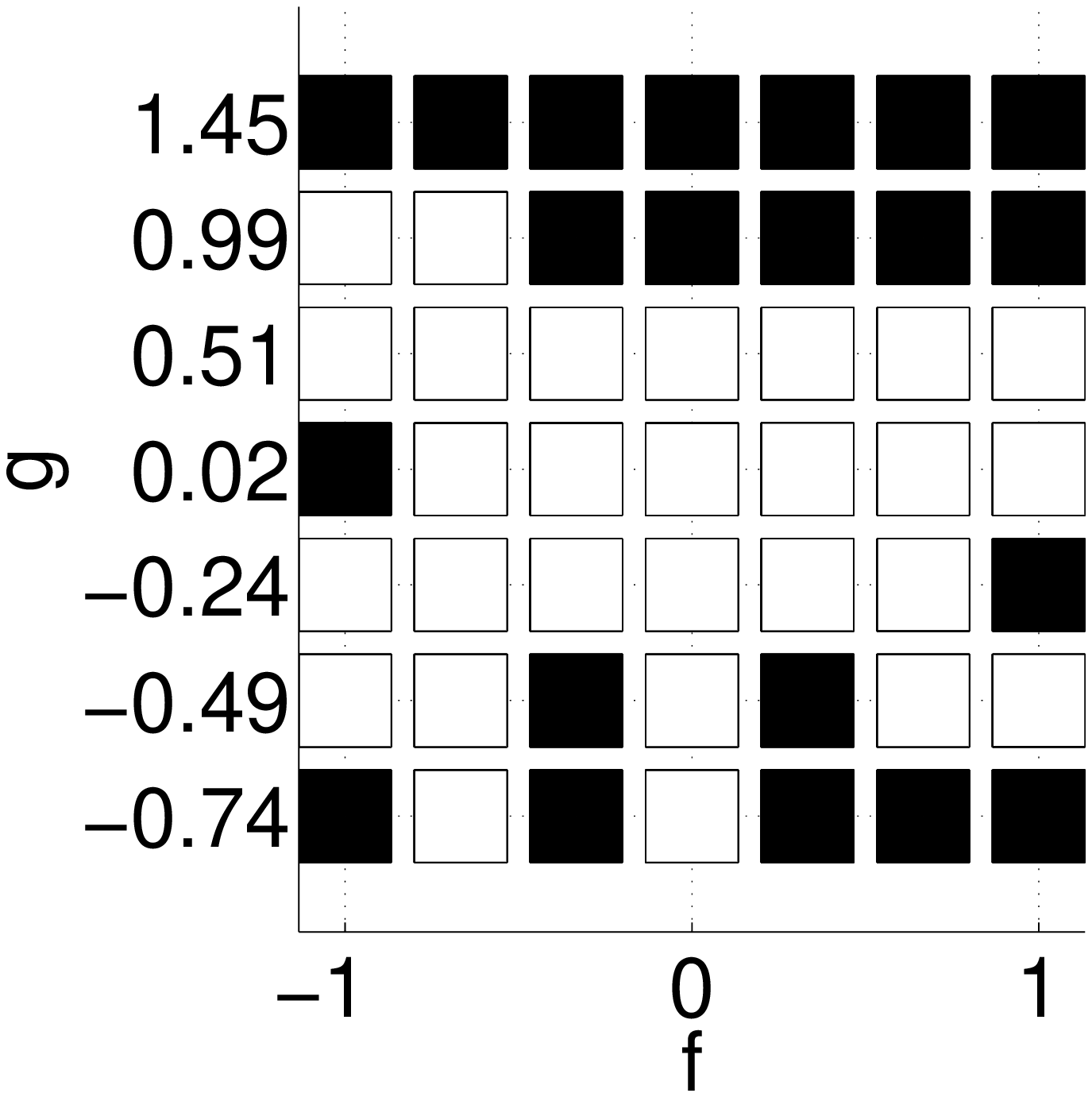}}
     {\includegraphics[width=.125\textwidth]{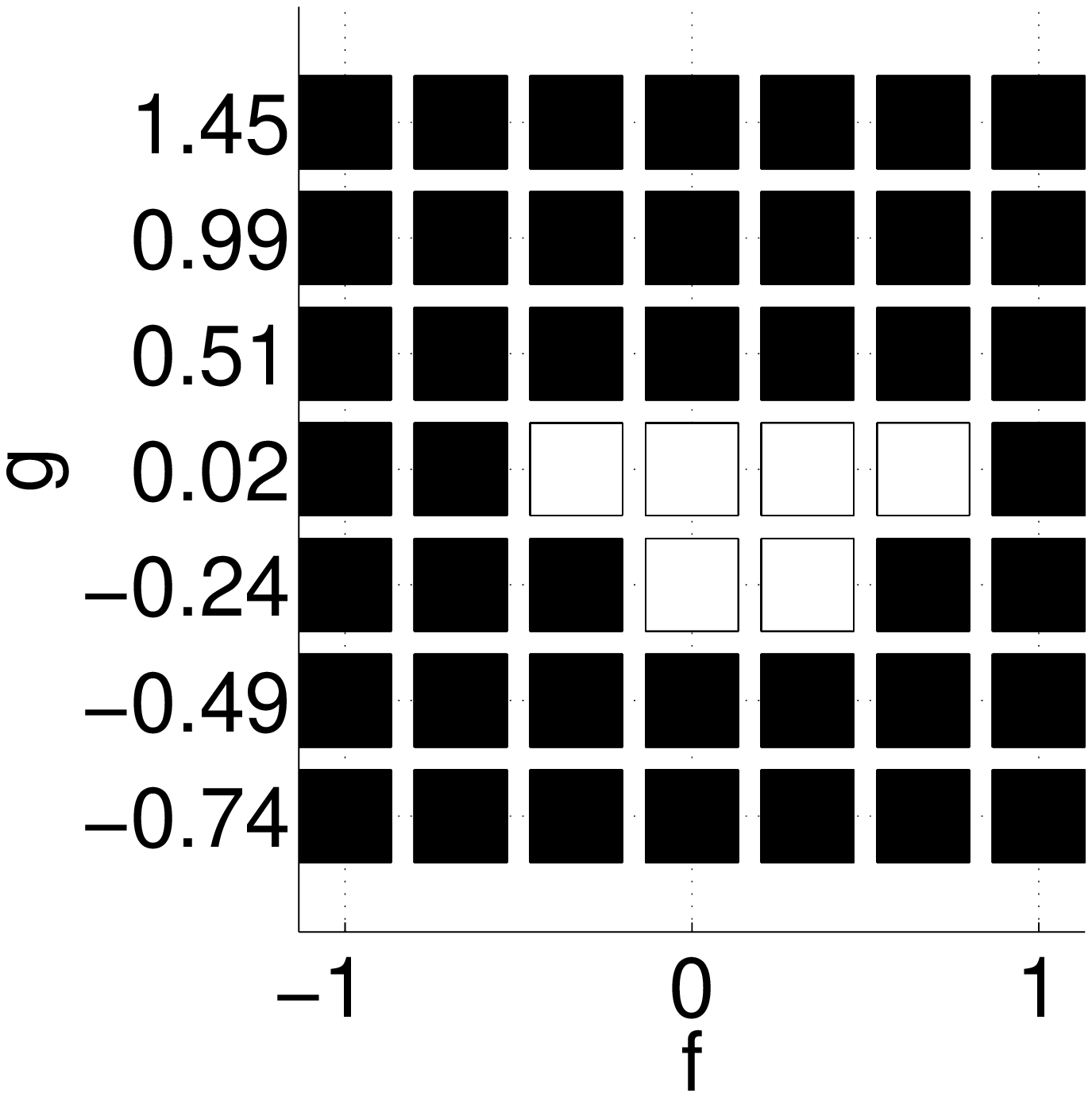}}
\caption{Upper panel: Experimental fidelity for the shaped
$\pi/2$-pulse (lhs) shown in Fig. \ref{fig:Ramsey_Fit} and of a
rectangular $\pi/2$-pulse (rhs) as a function of detuning $f$ and
amplitude $g$. The average statistical error of the measured
points for the shaped pulse is $\sigma_{av}=
 0.067 $, whereas $\sigma_{av}=
 0.086 $ for the rectangular pulse.
Lower panel: A white rectangle indicates $F/F_m > 0.90$ for the
shaped and the rectangular pulse, respectively.}
 \label{fig:645Res}
\end{figure}

The experimentally determined fidelity of this shaped pulse as a
function of $f$ and $g$ is displayed in Fig. \ref{fig:645Res}. For
reference, Fig. \ref{fig:645Res} also shows the experimental
fidelity obtained from a simple rectangular $\pi/2$ gate for the
same parameter range. For $f=0=g$ the simple pulse yields the
maximum fidelity $F_{m}= 0.899 \pm 0.065$, and the range of $f$ and
$g$ for which $F/F_m > 0.90$ is indicated in Fig. \ref{fig:645Res}
lower panel by white rectangles. The same maximum fidelity, $F_m=
0.900\pm 0.064$, is also reached using the shaped pulse. However,
this maximum value is maintained over a much wider parameter range
as is evident from Fig. \ref{fig:645Res}, thus demonstrating the
robustness of the shaped pulse against experimental errors and
intrinsic imperfections.

\begin{figure}
     {\includegraphics[width=.4\textwidth]{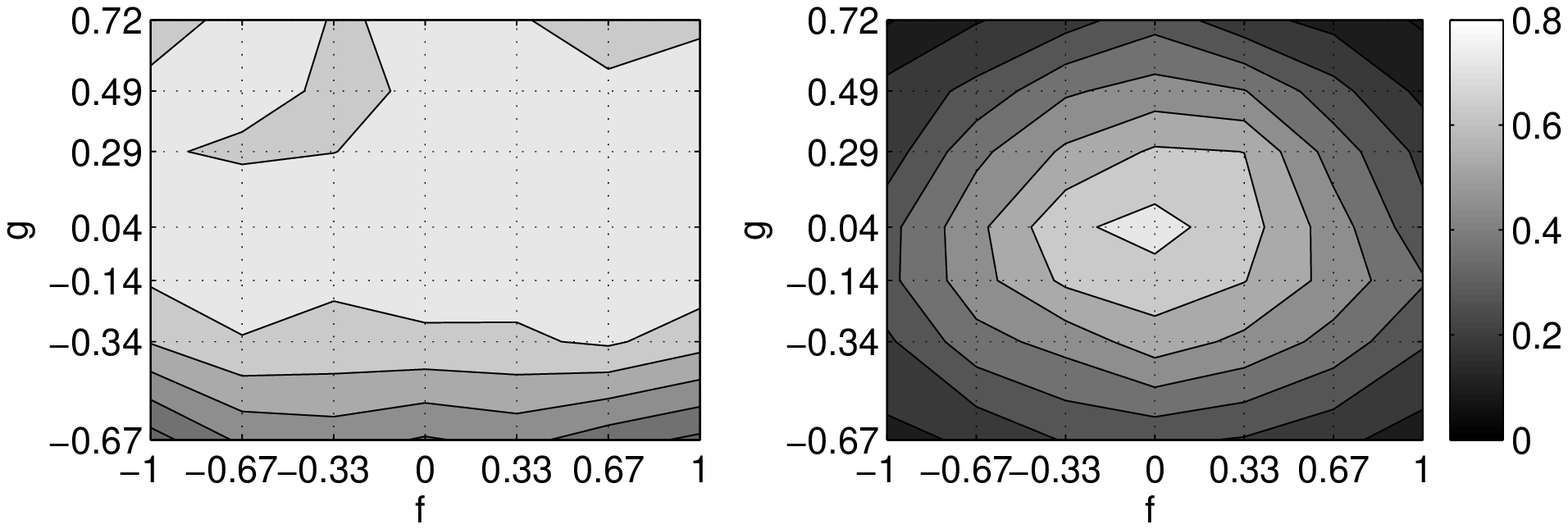}}
     {\includegraphics[width=.125\textwidth]{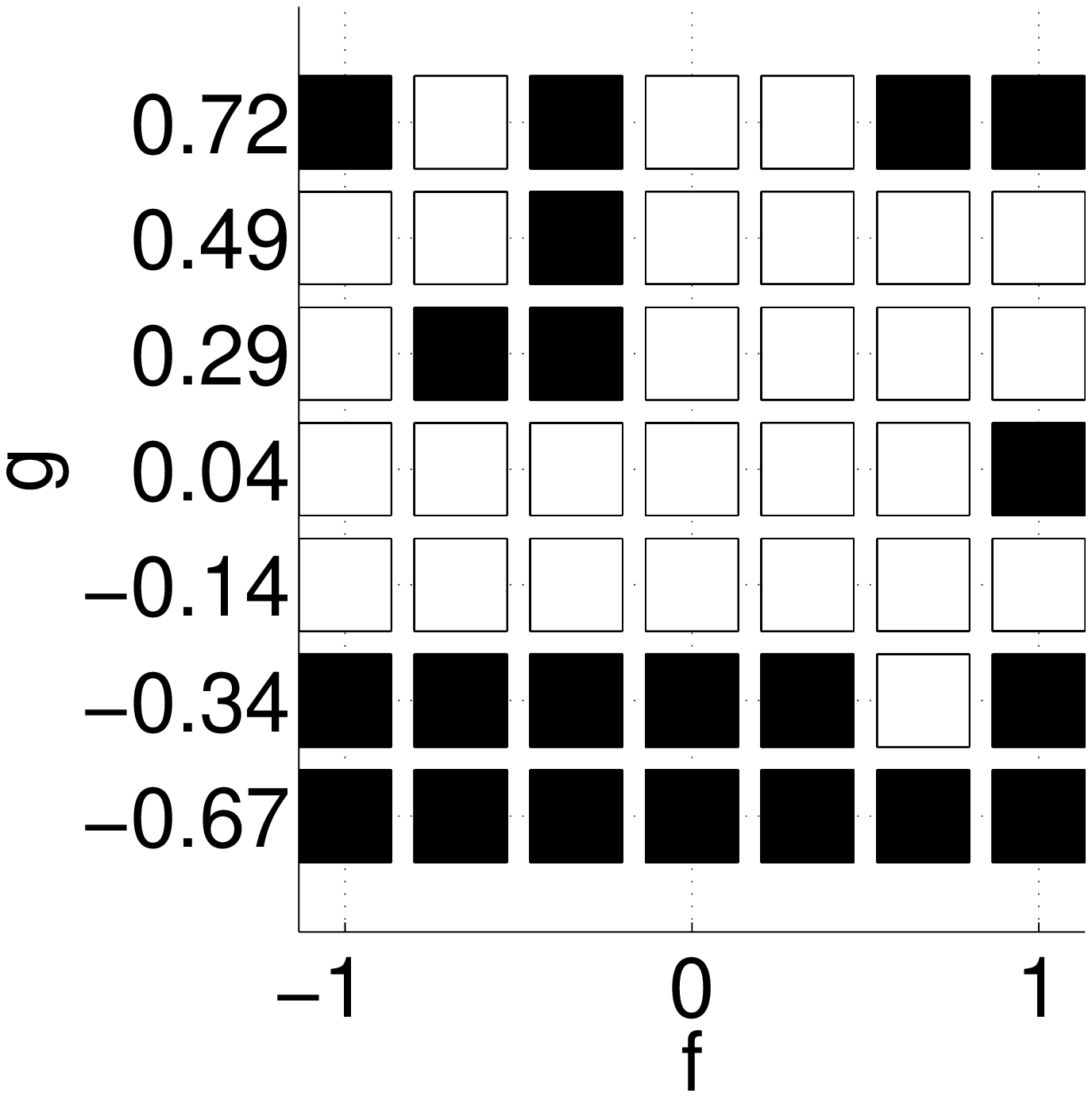}}
     {\includegraphics[width=.125\textwidth]{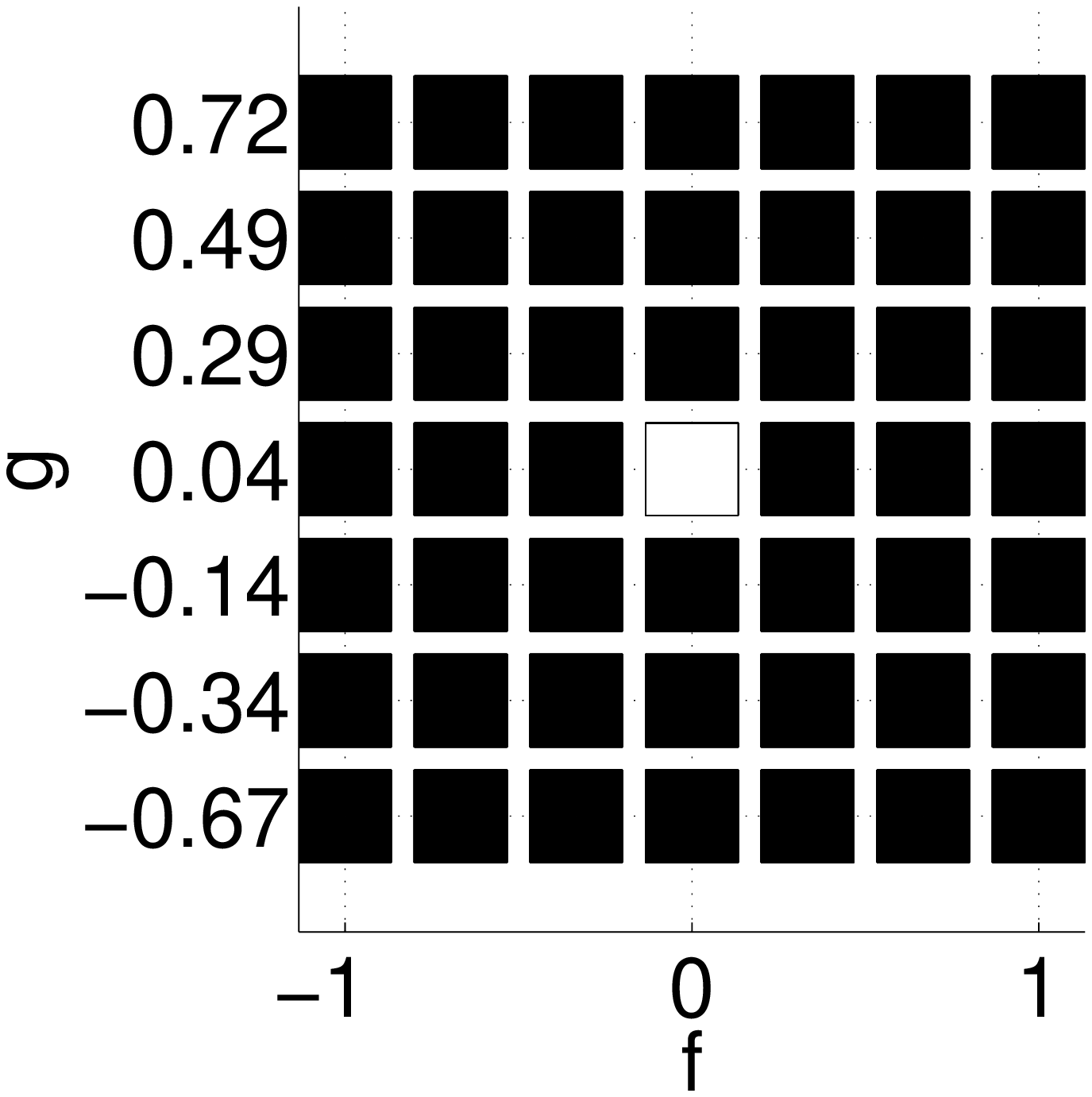}}
\caption{Upper panel: Experimental fidelity of a shaped
$\pi$-pulse (lhs, 445 steps, $\sigma_{av} = 0.028$), and the
corresponding rectangular $\pi$-pulse (rhs) as reference
($\sigma_{av} =0.020$). Lower panel: White rectangles show $F/F_m
> 0.96$.}
\label{fig:ShapedPi} 
\end{figure}

Fig. \ref{fig:ShapedPi} displays the experimental fidelity
obtained by using a shaped $\pi$-pulse consisting of 445 steps
with variable phase and amplitude subject to controlled errors.
Again, a rectangular pulse serves as an experimental reference
giving a maximum fidelity $F_m= 0.838 \pm 0.030$ for $f=0=g$ that
rapidly decreases for increasing $|f|$ or $|g|$. The shaped pulse,
in contrast, maintains the maximal possible fidelity over a wide
range of parameters as is evident in Fig. \ref{fig:ShapedPi}. A
shaped $\pi$-pulse consisting of 835 steps was also implemented
resulting in a parameter area indicating robustness against errors
that extends even further (not shown).

If one uses shaped pulses obtained from optimal control theory, even
with parameters that considerably deviate from their ideal values,
one still attains accurate performance of single qubit gates. Now,
we compare the performance of these pulses to some composite pulses
that have been devised to be effective against off-resonance,
amplitude, and pulse length errors. The resulting fidelity grids
were measured by sequentially performing the composite case and the
simple case for each point on the grid, and then repeating 300
times. In terms of the previously used description, sequence A and
sequence B are performed for each point on the grid, they are
followed by C and D, then the complete measurement procedure is
performed 300 times.

\begin{figure}
    \centering
    {\includegraphics[width=.4\textwidth]{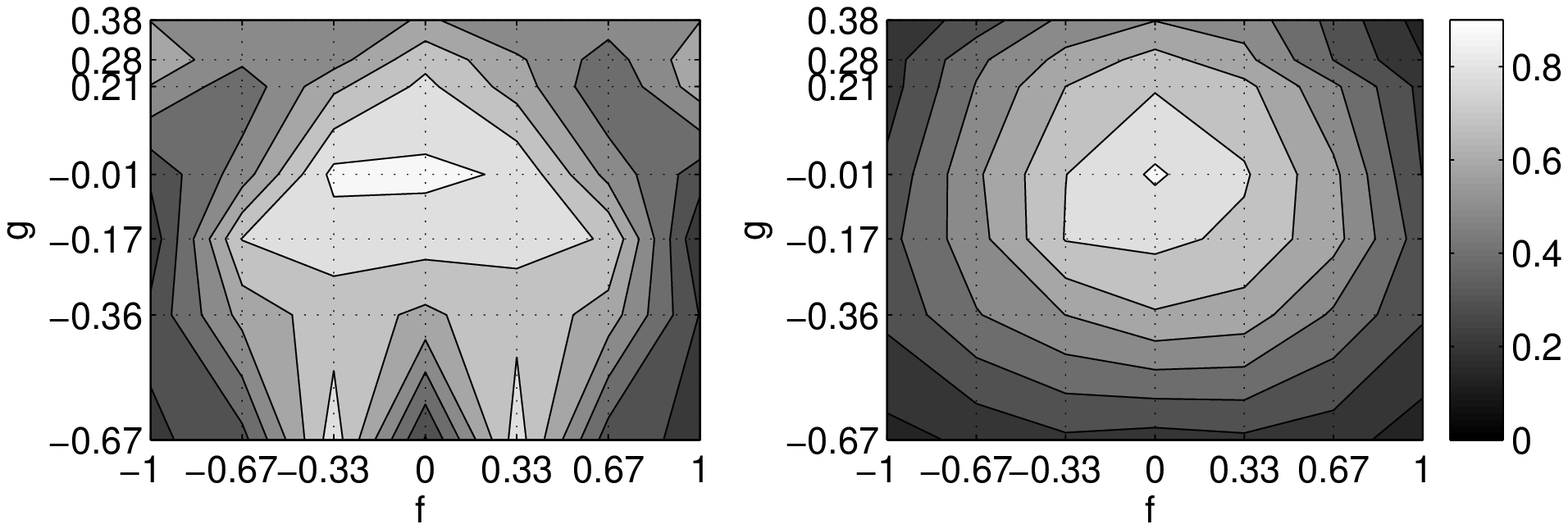}}
    {\includegraphics[width=.125\textwidth]{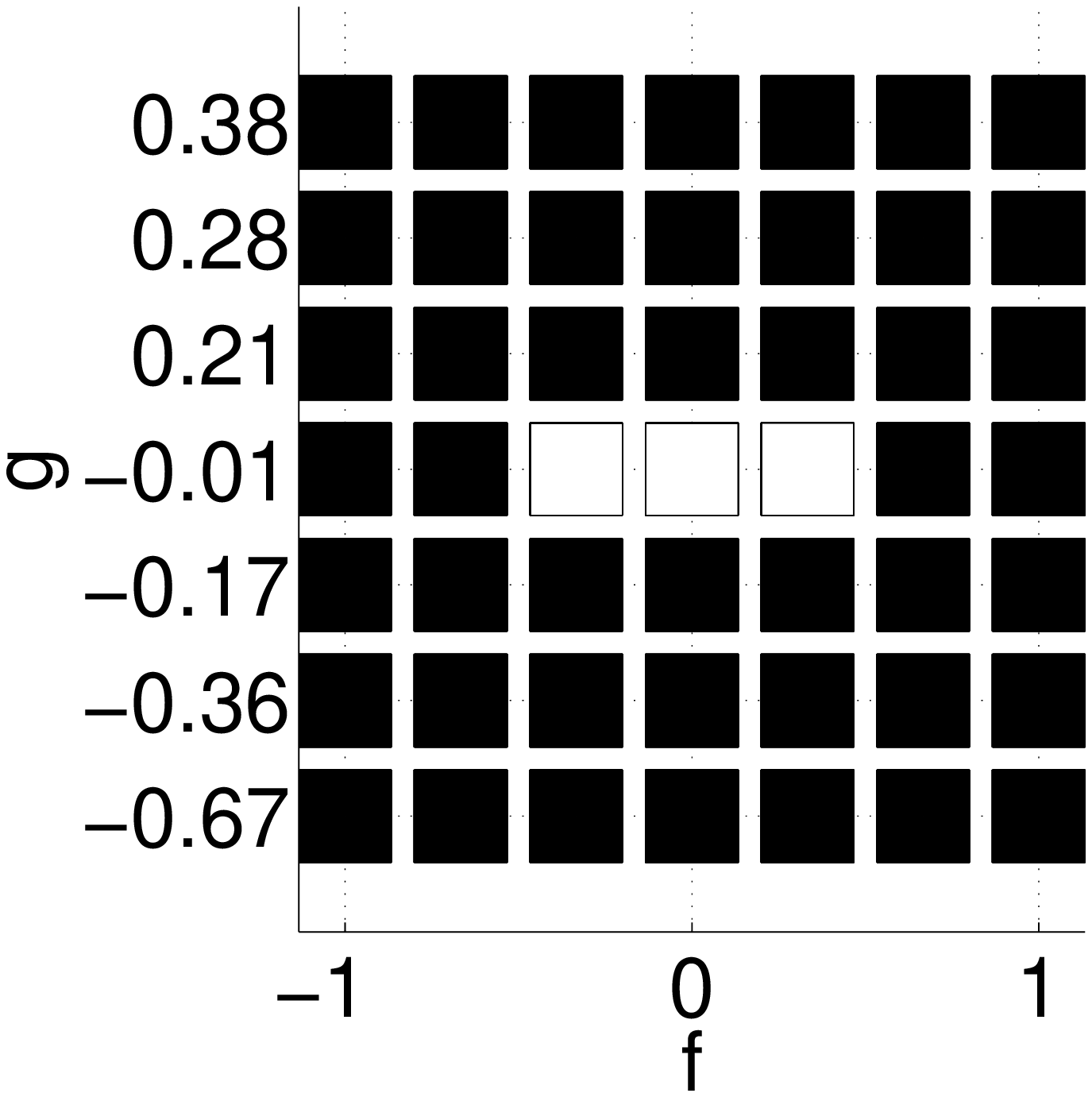}}
    {\includegraphics[width=.125\textwidth]{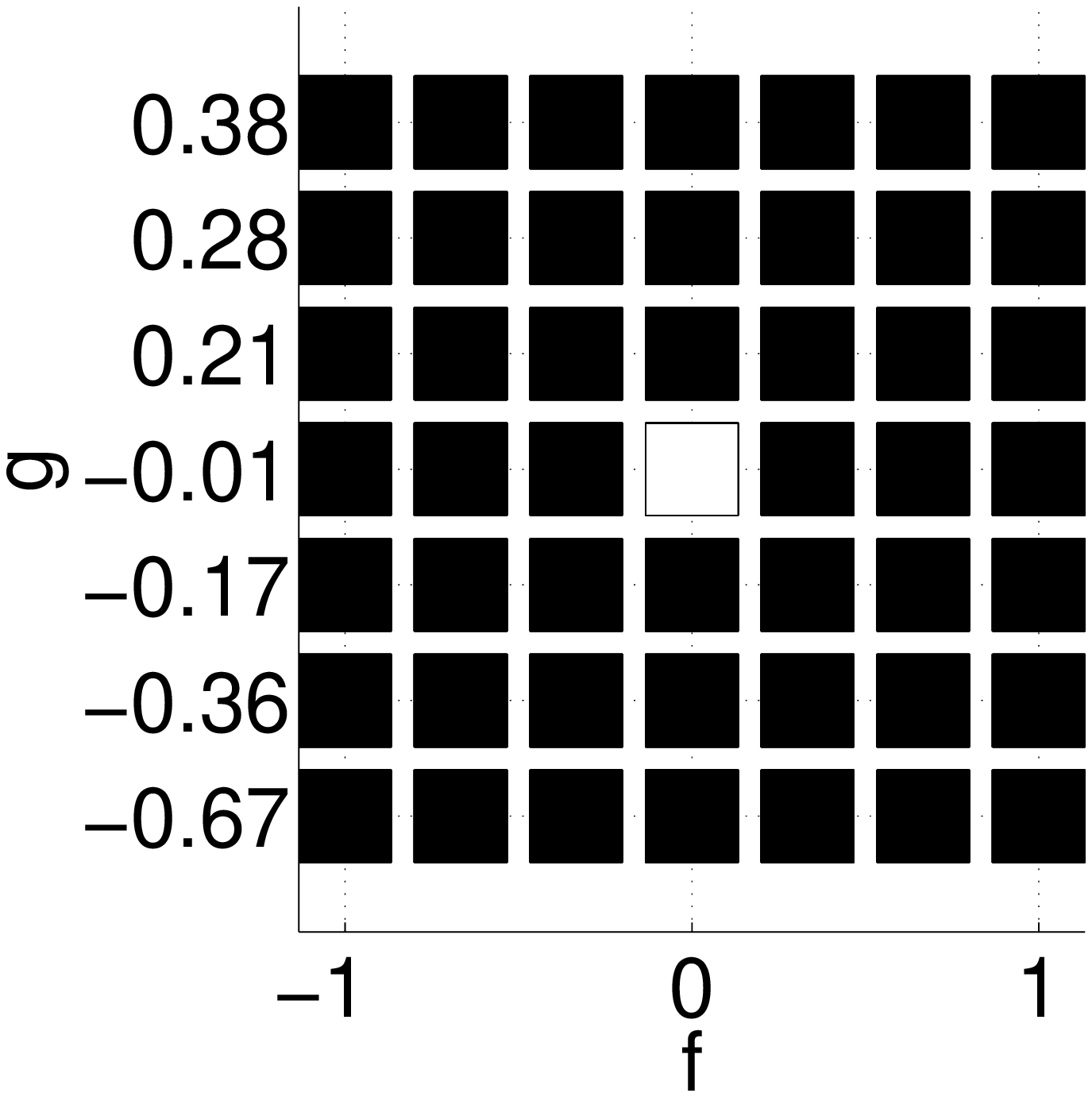}}
    {\includegraphics[width=.4\textwidth]{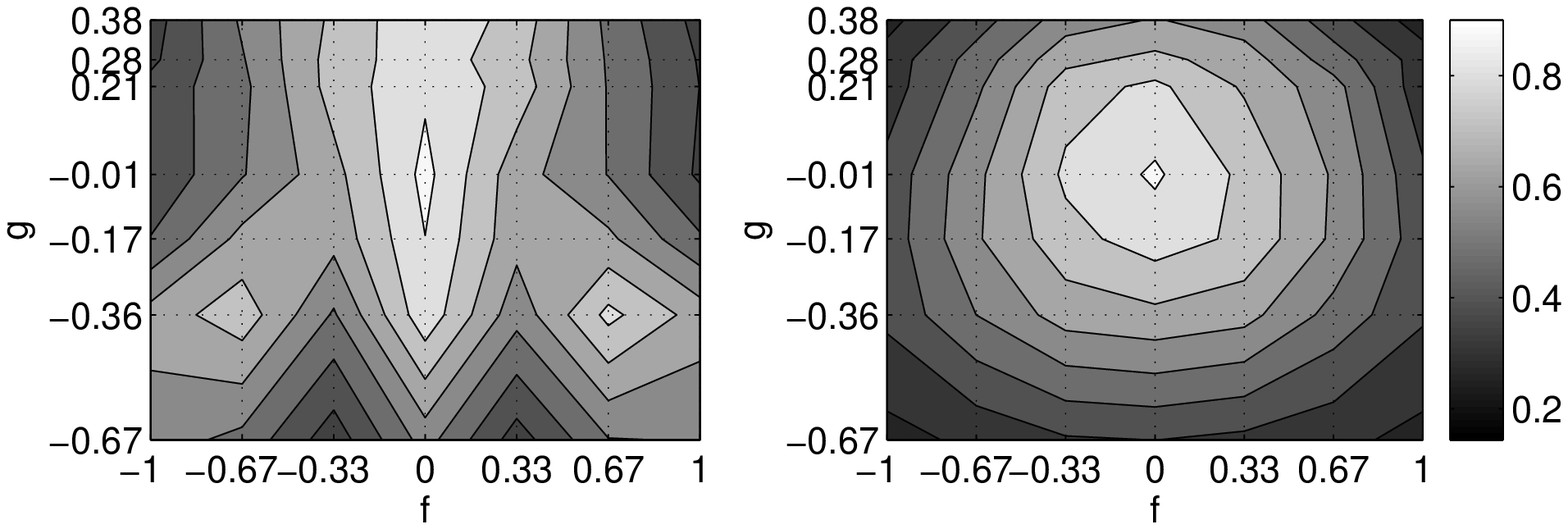}}
    {\includegraphics[width=.125\textwidth]{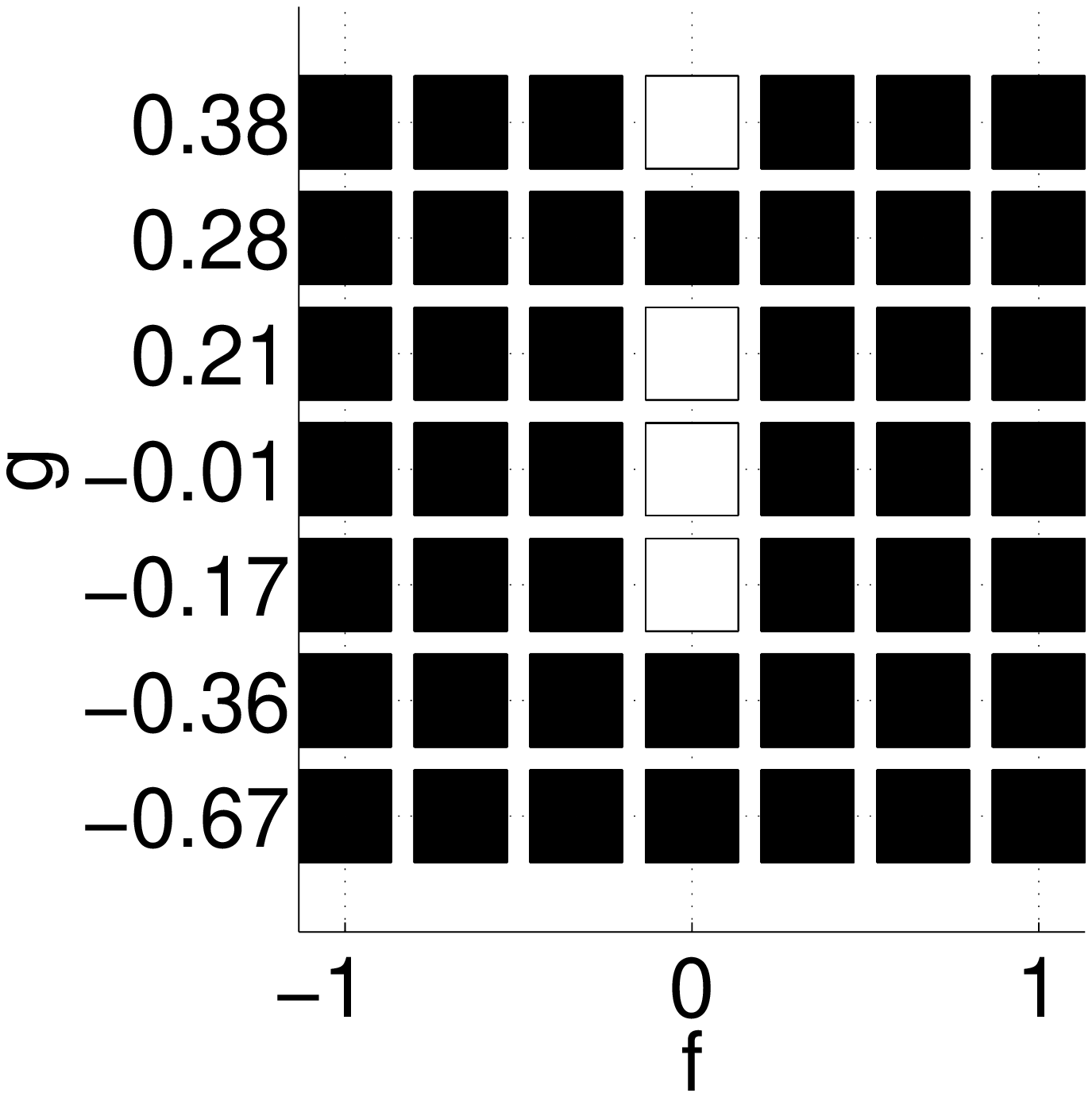}}
    {\includegraphics[width=.125\textwidth]{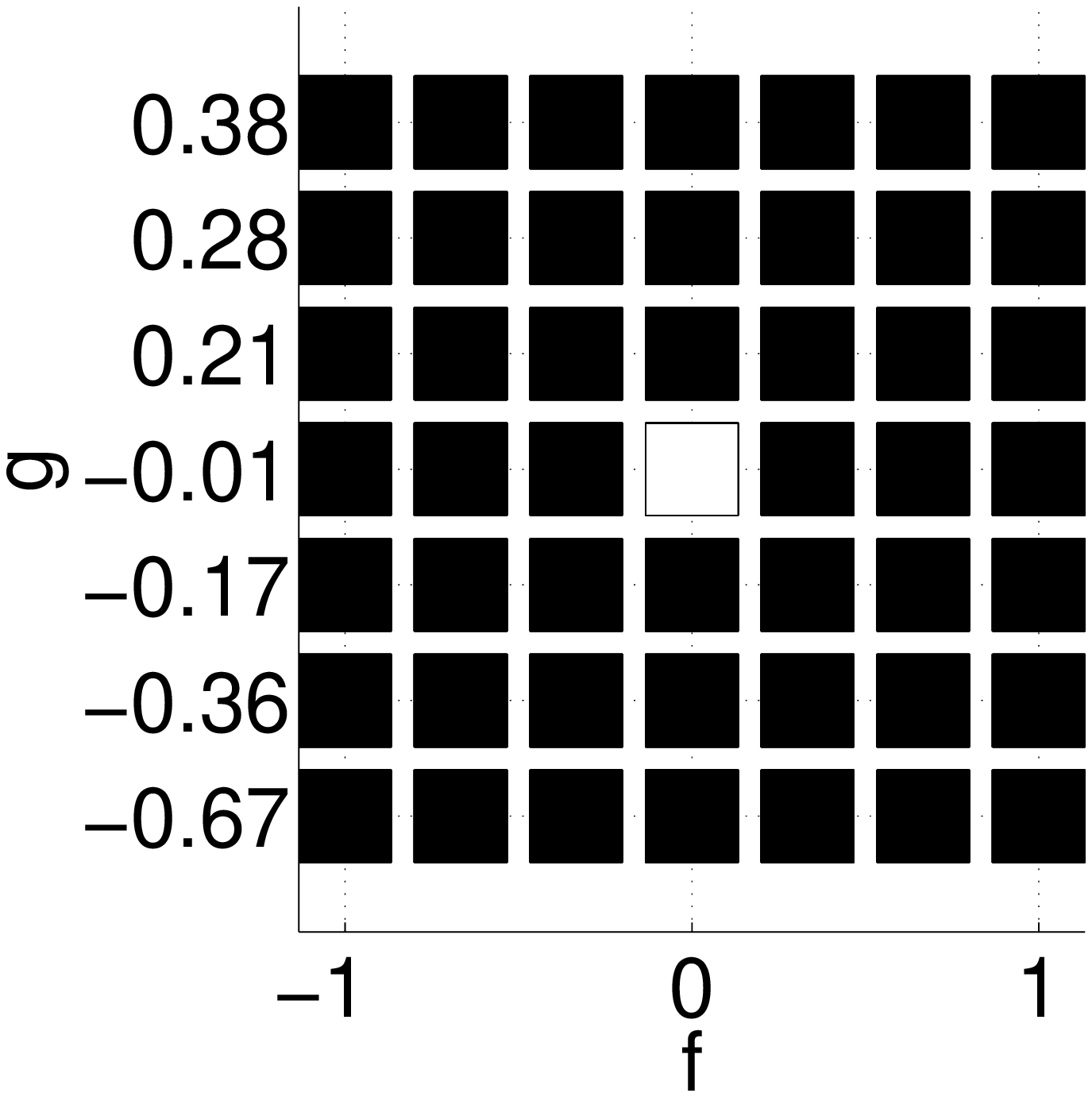}}
\caption{Upper panel: Measured fidelity for a CORPSE-type
composite pulse (lhs, $\theta = \pi$, $\sigma_{av} = 0.033$) with
the corresponding rectangular pulse for reference (rhs,
$\sigma_{av} = 0.030)$. Also shown are measured grids in terms of
$F/F_m
> 0.96$ where white is true and black is false. Lower panel: Measured fidelity
of a SCROFOLOUS-type composite pulse (lhs, $\theta = \pi$,
$\sigma_{av} = 0.0327$) with the corresponding rectangular pulse for
reference (rhs, $\sigma_{av} = 0.0299)$. White squares  in the
graphs below indicate $F/F_m
> 0.96$, $F_m= 0.915 \pm 0.043$.}
 \label{fig:CompPi}
\end{figure}

The composite pulse of type CORPSE(Compensation for Off-Resonance
with a Pulse Sequence) was derived with the aim of combatting
off-resonant errors \cite{Cummins03}. It consists of three pulses
where the first and the third pulses have equal phase values
$\Phi$ and the phase of the second pulse differs by $\pi$. The
nutation angles of the three pulses making up a $\pi$ composite
pulse are $420^\circ , 300^\circ , 60^\circ.$ Indeed, as shown in
Fig. \ref{fig:CompPi} upper panel the CORPSE $\pi$-pulse extends
the experimentally determined range of detuning $f$ over which a
fidelity $F/F_m > 0.96$ ($F_m =0.915 \pm 0.043$) is maintained as
compared to the rectangular pulse shown on the right of Fig.
\ref{fig:CompPi}. However, for compensating pulse area errors this
pulse sequence is less effective as is also evident from Fig.
\ref{fig:CompPi}.

The SCROFOLOUS (Short Composite ROtation For Undoing Length Over and
Under Shoot) pulse was derived with the aim of compensating for
pulse area errors. These pulses were derived by restricting the
phase and and nutation angles such that $\Phi_1 = \Phi_3 ; \theta_1
= \theta_3 $ with $\theta_2$ and $\Phi_2$ unbound. The angles are
chosen such that once again the fidelity is least effected by the
presence of errors \cite{Cummins03}. For a $\pi$-pulse these angles
are $\theta_1 =180^\circ, \theta_2 = 180^\circ; \Phi_1 = 60^\circ,
\Phi_2 = 300^\circ$. We show in Fig. \ref{fig:CompPi} lower panel
that errors in pulse area caused by power fluctuations are well
compensated for. On the other hand, detuning errors have the same
detrimental effect as is the case with a rectangular pulse.
Composite $\pi$ pulses, with g describing time errors rather than
amplitude errors were also implemented (not shown).

\begin{figure}
     {\includegraphics[width=.4\textwidth]{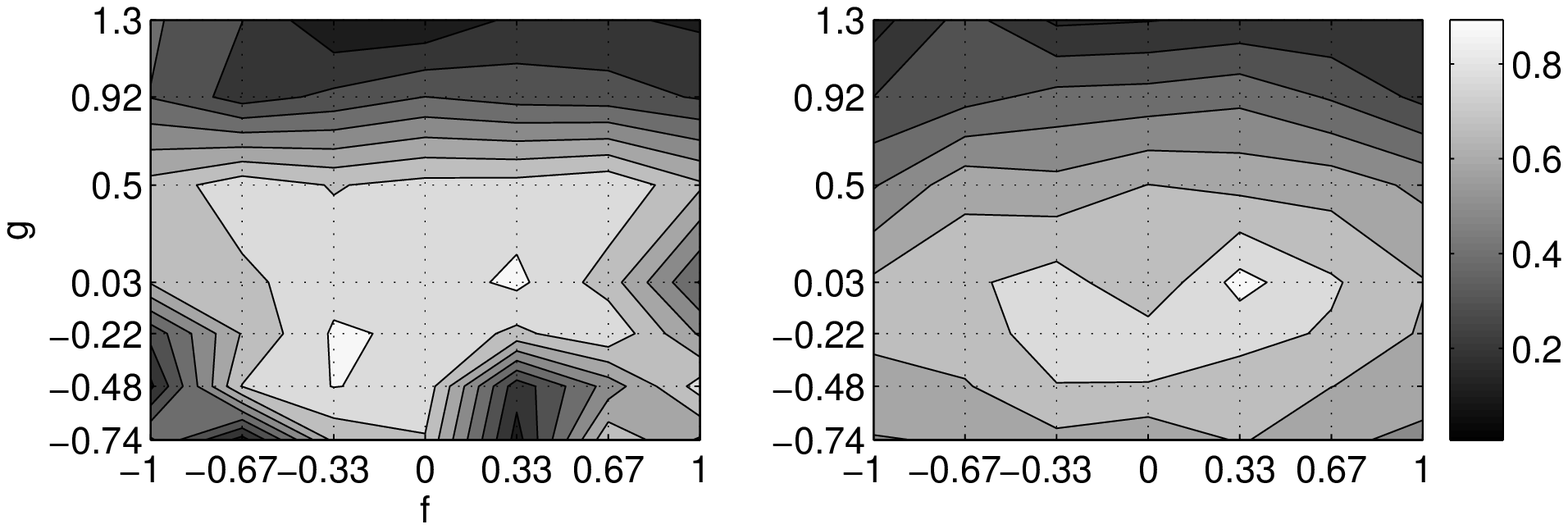}}
     {\includegraphics[width=.125\textwidth]{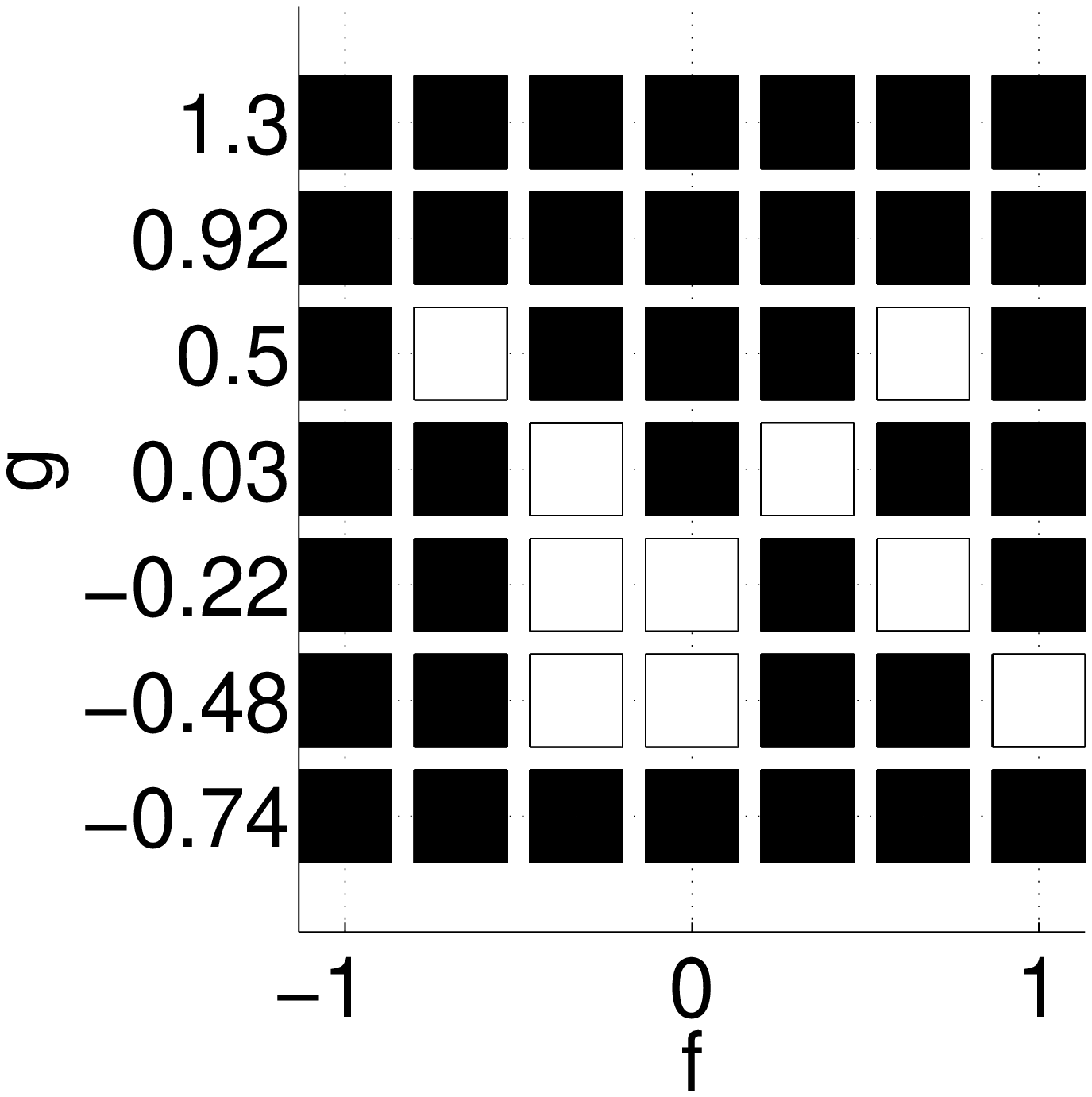}}
     {\includegraphics[width=.125\textwidth]{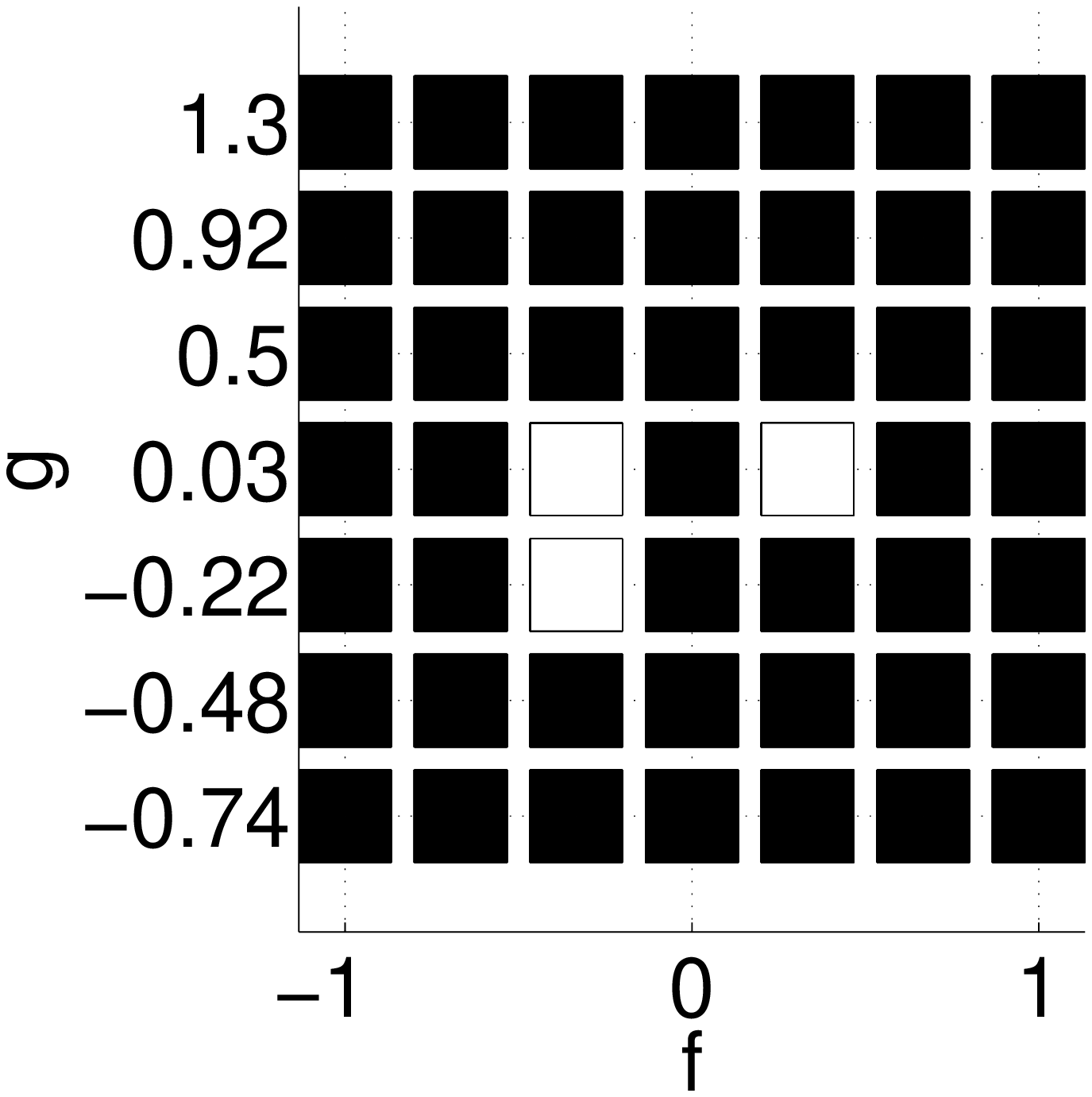}}
      \caption{Upper panel: Experimental Fidelity for the BB1
$(\frac{\theta}{2})-W-R(\frac{\theta}{2})$ sequence (lhs), where
$\theta = \frac{\pi}{2}$ (left, $\sigma_{av}= 0.072$) and for a
corresponding rectangular pulse (rhs). Lower panel: Measured grids
in terms of $F/F_m
> 0.90, F_m = 0.936 \pm 0.052$}
\label{fig:BB1RWRPi_2Res}
\end{figure}

The BB1 (Broadband) composite pulse comprises a sequence $W =
180_{\Phi 1} -360_{\Phi 2}-180_{\Phi 3}$. When the desired
rotation is R($\theta$), the complete rotation can be performed as
$R(\theta)-W,W-R(\theta)$ or $(\theta/2)-W-R(\theta/2)$. The
measurement procedure follows the procedure outlined above for the
shaped $\pi/2$-pulse. Shown in Fig. \ref{fig:BB1RWRPi_2Res} are
the results of using the BB1 composite pulse
$R(\theta/2)-W-R(\theta/2)$ with $\theta=\pi/2$. The experimental
data show that BB1 is effective for compensation of errors in both
detuning and pulse area. The parameter range over which error
resistant pulses are effective is greater than that of a simple
pulse but is more restricted than with the shaped pulse shown in
Fig. \ref{fig:645Res}.

A comparison of the performance of shaped pulses developed using
optimal control theory with simple rectangular pulses or composite
pulses reveals an evident advantage of these shaped pulses in
terms of robustness against experimental errors and
indeterministic system parameters, while the lengths of both types
of pulses are comparable (here, with $\Omega= 2\pi\times 10$ kHz,
for instance the pulse in Fig. \ref{fig:ShapedPi} takes 223 $\mu$s
compared to 217 $\mu$s for a CORPSE pulse of Fig.
\ref{fig:CompPi}). This will make shaped pulses based on optimal
control theory an important tool in order to achieve quantum gates
with trapped ions with low error probability and thus come a step
closer to fault-tolerant quantum computing.

We acknowledge discussions with S. Glaser, T. Schulte-Herbrueggen,
C. Wei\ss,  who also supplied simulation software, and support from
the European Union IP QAP.


\begin{references}
\bibitem{Aliferis06} P. Aliferis, D. Gottesman, and J. Preskill,
Quant. Inf. Comp. {\bf 6}, 97 (2006).
\bibitem{Knill06} E. Knill, Nature {\bf 434}, 39 (2005).
\bibitem{Barenco95}A. Barenco et al., Phys. Rev. A, {\bf 52}, 3457
(1995); For certain quantum computational tasks, gates that
simultaneously act on more than two qubits have been shown to be
more efficient than the use of two-qubit gates.
\bibitem{Leibfried05}D. Leibfried et al., Nature {\bf 438}, 639
(2005); H. H\"{a}ffner et al., {\it ibid.} p. 643 (2005); P.C.
Haljan et al., Phys. Rev. A {\bf 72}, 062316 (2005); J.P. Home et
al., New J. Phys. {\bf 8}, 188 (2006).
\bibitem{SchmidtKaler03}F. Schmidt-Kaler et al., Nature {\bf 422},
408 (2003).
\bibitem{Leibfried03}D. Leibfried et al., Nature {\bf 422}, 412
(2003).
\bibitem{Kielpinski02}D. Kielpinski, C. Monroe, D. J. Wineland, Nature {\bf 417},
709 (2002).
\bibitem{Wunderlich02}Chr. Wunderlich,  in {\em Laser
Physics at the Limit} (Springer, Heidelberg, 2002), p. 261;
available as quant-ph/0111158; F. Mintert, Chr. Wunderlich, Phys.
Rev. Lett. {\bf 87},  257904 (2001); D. Mc Hugh, J. Twamley, Phys.
Rev. A {\bf 71}, 012315 (2005).
\bibitem{Skinner03} T.E.~Skinner, T.O.~Reiss, B. Luy, N. Khaneja and S.~Glaser,
 J. Mag. Res. {\bf 163}, 8 (2003).
\bibitem{Cummins03} H.~Cummins, G.~Llewellyn, and J.~Jones, Phys. Rev. A {\bf
67}, 042308 (2003).
\bibitem{Luy04} K.~Kobzar, T.E.~Skinner,N.Khaneja, S. Glaser and B.~Luy, J. Mag. Res. {\bf
170}, 8, (2003).
\bibitem{Wunderlich03}
Chr. Wunderlich and Chr. Balzer, Adv. At. Mol. Opt. Phys. {\bf 49},
293 (2003).

\end{references}
\end{document}